\documentclass[twocolumn,tighten]{aastex62}
\usepackage{natbib}
\usepackage{amsmath}
\usepackage{listings}
\usepackage[normalem]{ulem}
\usepackage[section]{placeins}
\usepackage{color}

\received{2018 June 28}
\revised{2018 August 10}
\accepted{2018 August 14}

\newcommand{\comments}[1]{} 

\shorttitle{Formation of LISA Black Hole Binaries in Merging Dwarf Galaxies}
\shortauthors{T.~Tamfal et al.}

\begin{document}

\title{Formation of LISA Black Hole Binaries in Merging Dwarf Galaxies: the Imprint of Dark Matter}


\correspondingauthor{Tomas Tamfal}
\email{tomas.tamfal@uzh.ch}

\author[0000-0003-1773-9349]{Tomas Tamfal}
\affiliation{Center for Theoretical Astrophysics and Cosmology, Institute for Computational Science, University of Zurich, Winterthurerstrasse 190, CH-8057 Z\"urich, Switzerland}

\author[0000-0002-1786-963X]{Pedro R. Capelo}
\affiliation{Center for Theoretical Astrophysics and Cosmology, Institute for Computational Science, University of Zurich, Winterthurerstrasse 190, CH-8057 Z\"urich, Switzerland}

\author{Stelios Kazantzidis}
\affiliation{Section of Astrophysics, Astronomy and Mechanics, Department of Physics, National and Kapodistrian University of Athens, 15784 Zografos, Athens, Greece}

\author[0000-0002-7078-2074]{Lucio Mayer}
\affiliation{Center for Theoretical Astrophysics and Cosmology, Institute for Computational Science, University of Zurich, Winterthurerstrasse 190, CH-8057 Z\"urich, Switzerland}

\author[0000-0002-0757-5195]{Douglas Potter}
\affiliation{Center for Theoretical Astrophysics and Cosmology, Institute for Computational Science, University of Zurich, Winterthurerstrasse 190, CH-8057 Z\"urich, Switzerland}

\author[0000-0001-7565-8622]{Joachim Stadel}
\affiliation{Center for Theoretical Astrophysics and Cosmology, Institute for Computational Science, University of Zurich, Winterthurerstrasse 190, CH-8057 Z\"urich, Switzerland}

\author{Lawrence M. Widrow}
\affiliation{Department of Physics, Engineering Physics $\&$ Astronomy, Queen's University, Stirling Hall, Kingston, ON K7L 3N6, Canada}


\begin{abstract}
Theoretical models for the expected merger rates of intermediate-mass black holes (IMBHs) are vital for planned gravitational-wave detection experiments such as the Laser Interferometer Space Antenna (LISA). Using collisionless $N$-body simulations of dwarf galaxy (DG) mergers, we examine how the orbital decay of IMBHs and the efficiency of IMBH binary formation depend on the central dark matter (DM) density profile of the merging DGs. Specifically, we explore various asymptotic inner slopes $\gamma$ of the DG's DM density distribution, ranging from steep cusps ($\gamma=1$) to shallower density profiles ($\gamma<1$), motivated by well-known baryonic-feedback effects as well as by DM models that differ from cold DM at the scales of DGs. We find that the inner DM slope is crucial for the formation (or lack thereof) of an IMBH binary; only mergers between DGs with cuspy DM profiles ($\gamma=1$) are favourable to forming a hard IMBH binary, whereas when $\gamma<1$ the IMBHs stall at a separation of 50--100~pc. Consequently, the rate of LISA signals from IMBH coalescence will be determined by the fraction of DGs with a cuspy DM profile. Conversely, the LISA event rates at IMBH mass scales offer in principle a novel way to place constraints on the inner structure of DM halos in DGs and address the core--cusp controversy. We also show that, with spatial resolutions of $\sim$0.1~kpc, as often adopted in cosmological simulations, all IMBHs stall, independent of $\gamma$. This suggests caution in employing cosmological simulations of galaxy formation to study BH dynamics in DGs.
\end{abstract}



\section{Introduction}\label{sec:Introduction}

The gravitational-wave (GW) signal from merging intermediate-mass black holes (IMBHs) in the mass range $10^4$--$10^6$~M$_{\odot}$ is one of the best anticipated targets of the Lisa Interferometer Space Antenna \citep[LISA;][]{Amaro-Seoane_et_al_2017}. Given their relatively low mass, such BHs have been linked to \textcolor{black}{globular clusters \citep[within which they can merge with stellar-mass BHs and produce LISA events; e.g.][]{Fragione_et_al_2018a,Fragione_et_al_2018b}, massive star clusters \citep[where they can be detected via tidal disruption events; e.g.][]{Lin_et_al_2018}, and} dwarf galaxies (DGs). DGs are extremely numerous and experience on average three major mergers in their lifetime \citep[][]{Fakhouri_et_al_2010}, therefore possibly providing several LISA signals. Recently, observational work has revealed a population of IMBHs through the discovery of low-luminosity active galactic nuclei (AGN) in DGs both in the local Universe \citep[e.g.][]{Reines_et_al_2013,Baldassare_et_al_2015} and at high redshift \citep[$z \lesssim 2.4$;][]{Mezcua_et_al_2018}. The best known example with multi-wavelength observations is RGG~118, hosting an IMBH with a mass of $\sim$$5 \times 10^4$~M$_{\odot}$ \citep[][]{Baldassare_et_al_2015}.

Providing theoretical predictions for the expected merger rates of IMBHs is thus highly relevant to LISA. Recently, a number of studies have attempted to address the evolution of IMBHs in merging DGs utilizing cosmological simulations \textcolor{black}{\citep[][]{Habouzit_et_al_2017,Tremmel_et_al_2015,Bellovary_et_al_2018}}. However, due to limited resolution, such contributions are unable to resolve not only the formation of the IMBH binary, which would require pc-scale resolution \citep[][]{Mayer_et_al_2007,Pfister_et_al_2017}, but also the preceding pairing phase in the merger remnant \citep[][]{Chapon_et_al_2013,Pfister_et_al_2017}.

The dark matter (DM) distribution in the central regions of DGs may significantly affect the orbital decay of IMBHs and the efficiency of IMBH binary formation in interacting DGs. This is because dynamical friction (DF) depends on the background mass distribution \citep[][]{Colpi_2014} which, in the case of DGs, can be dominated by DM even at small radii. In particular, DF in constant-density cores is known to lead to the stalling of sinking perturbers (\citealt{Goerdt_et_al_2006,Cole_et_al_2012,Petts_et_al_2016}; see also \citealt{DiCintio_et_al_2017}). Therefore, whether or not the DM in DGs is described according to the Navarro--Frenk--White \citep[NFW;][]{Navarro_et_al_1996} profile, expected in cold-DM (CDM) cosmologies, or follows shallower mass distributions \citep[e.g.][]{Weinberg_et_al_2015,Read_et_al_2016} becomes particularly relevant. Shallow DM density profiles can arise either due to galaxy formation processes \citep[e.g. baryonic outflows triggered by supernova explosions; e.g.][]{Governato_et_al_2010}, or as a consequence of modifications of the underlying DM model (e.g. self-interacting DM, \citealt{Spergel_Steinhardt_2000}; or fuzzy DM, \citealt{Hui_et_al_2017}). Currently, it is at least clear that a wide diversity of DM distributions exists in DGs \citep[][]{Oman_et_al_2017}, and that in some cases models with nearly constant density DM cores appear to reproduce observed galaxy rotation curves better \citep[e.g.][]{Brooks_et_al_2017}.

There has been no quantitative work aimed at elucidating the effect of the DM density profile on the orbital decay of IMBHs and the efficiency of IMBH binary formation in merging DGs. Here we investigate this novel aspect of the co-evolution of galaxies and their central BHs via a series of high-resolution controlled merger simulations of DGs embedded in DM halos with different density distributions.


\section{Numerical Setup}\label{sec:Numerical_setup}


\subsection{Initial conditions}\label{sec:Initial_conditions}

We employed the methods described in \citet{Kuijken_Dubinski_1995}, \citet{Widrow_Dubinski_2005}, and \citet{Widrow_et_al_2008} to generate self-consistent $N$-body models of axisymmetric DGs consisting of exponential stellar disks, central BHs, and extended DM halos whose density profiles followed the general form \citep[e.g.][]{Lokas_2002}

\begin{equation}
\rho(r) = \frac{\rho_{\rm s}} {(r/r_{\rm s})^\gamma (1 + r/r_{\rm s})^{3-\gamma}},
\end{equation}

\noindent where $\gamma$ and $r_{\rm s}$ denote the asymptotic inner slope and the scale radius of the profile, respectively. The characteristic inner density, $\rho_{\rm s}$, depends on $\gamma$, the halo formation epoch, and the present-day values of the cosmological parameters (we adopt the concordance $\Lambda$-CDM cosmogony and $z = 0$).

\begin{deluxetable}{cccc}[htb!]
\tablecaption{Particle specifications} \tablecolumns{4}
\tablehead{\colhead{Resolution:}& high & low}
\startdata
$N_{\rm DM}$, $N_{\star}$, $N_{\rm BH}$ & $5 \times 10^7$, $10^7$, 1 & $10^6$, $5 \times 10^5$, 1 \\
$m_{\rm DM}$, $m_{\star}$, $M_{\rm BH}$ & 400, 40, $10^5$ & $2 \times 10^4$, 800, $10^5$ \\
$\epsilon_{\rm DM}$, $\epsilon_{\star}$, $\epsilon_{\rm BH}$ & 4, 1, 1 & 326, 109, 109 \\
\enddata
\tablecomments{From top to bottom: particle number, mass (in M$_{\odot}$), and gravitational softening (in pc) of DM, stellar, and BH particles per galaxy.}
\label{tab:simulations}
\vspace{-30pt}
\end{deluxetable}

\begin{figure*}[htb!]
\includegraphics[width=0.48\textwidth]{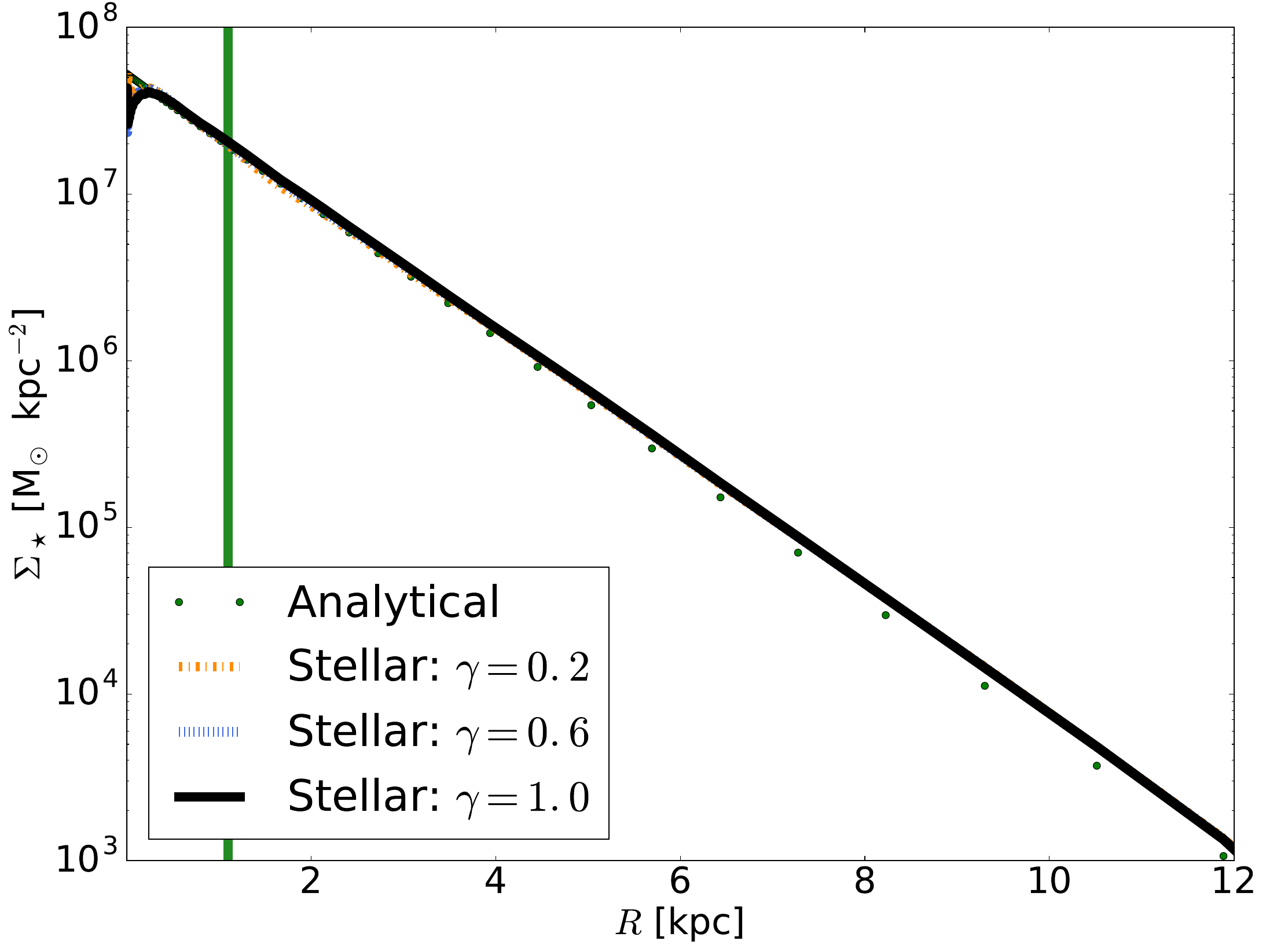}
\includegraphics[width=0.48\textwidth]{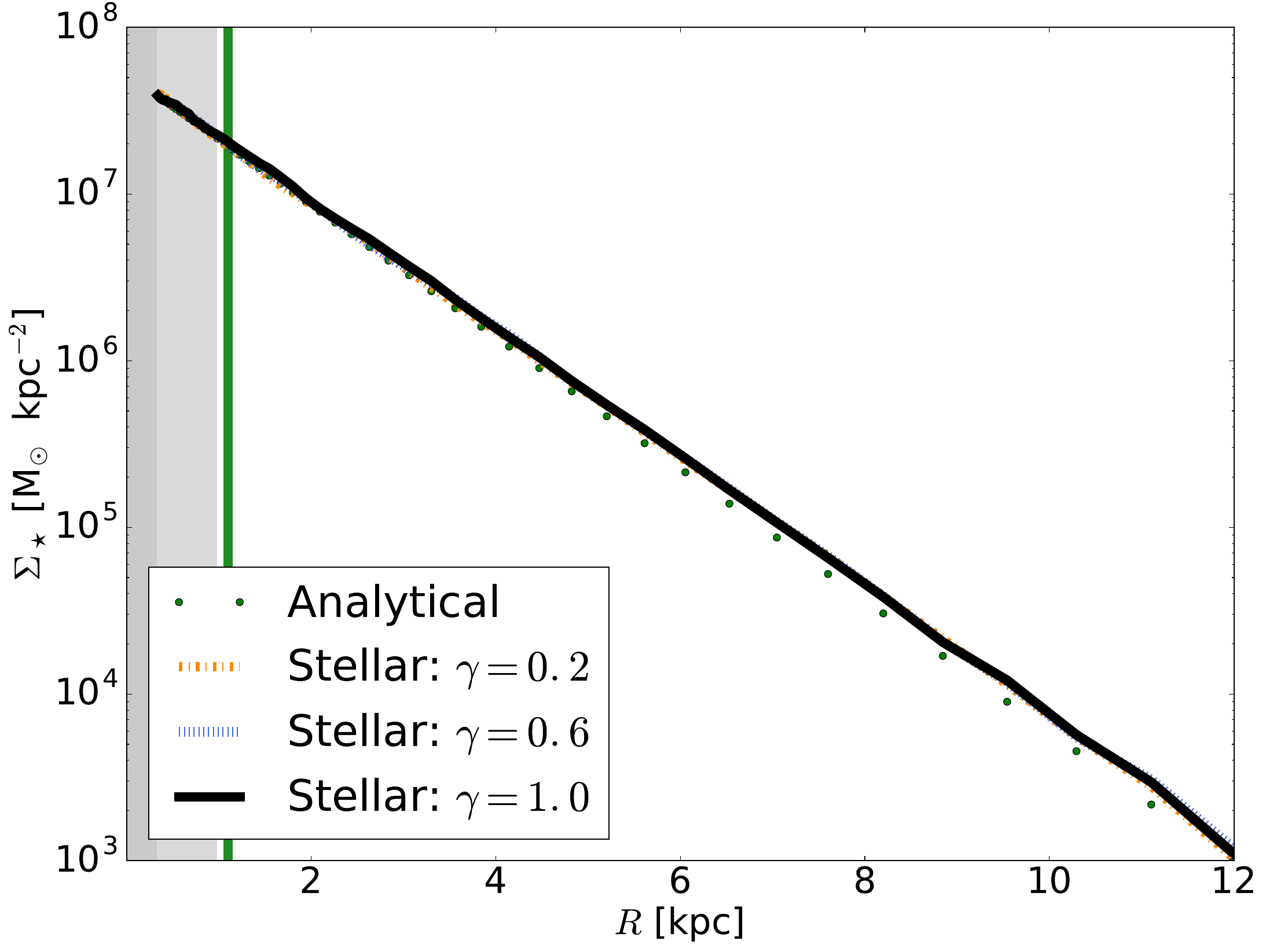}\\
\includegraphics[width=0.48\textwidth]{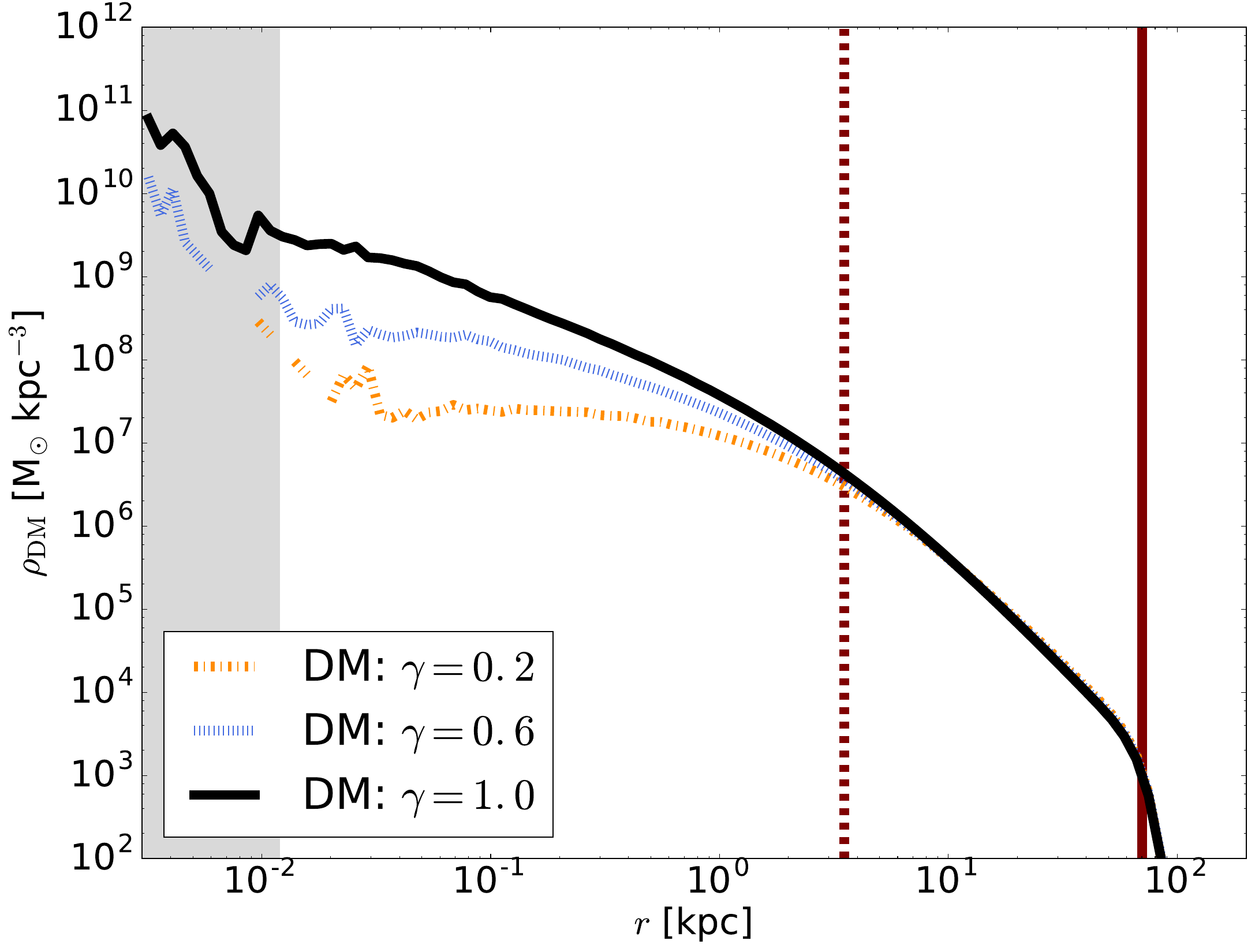}
\includegraphics[width=0.48\textwidth]{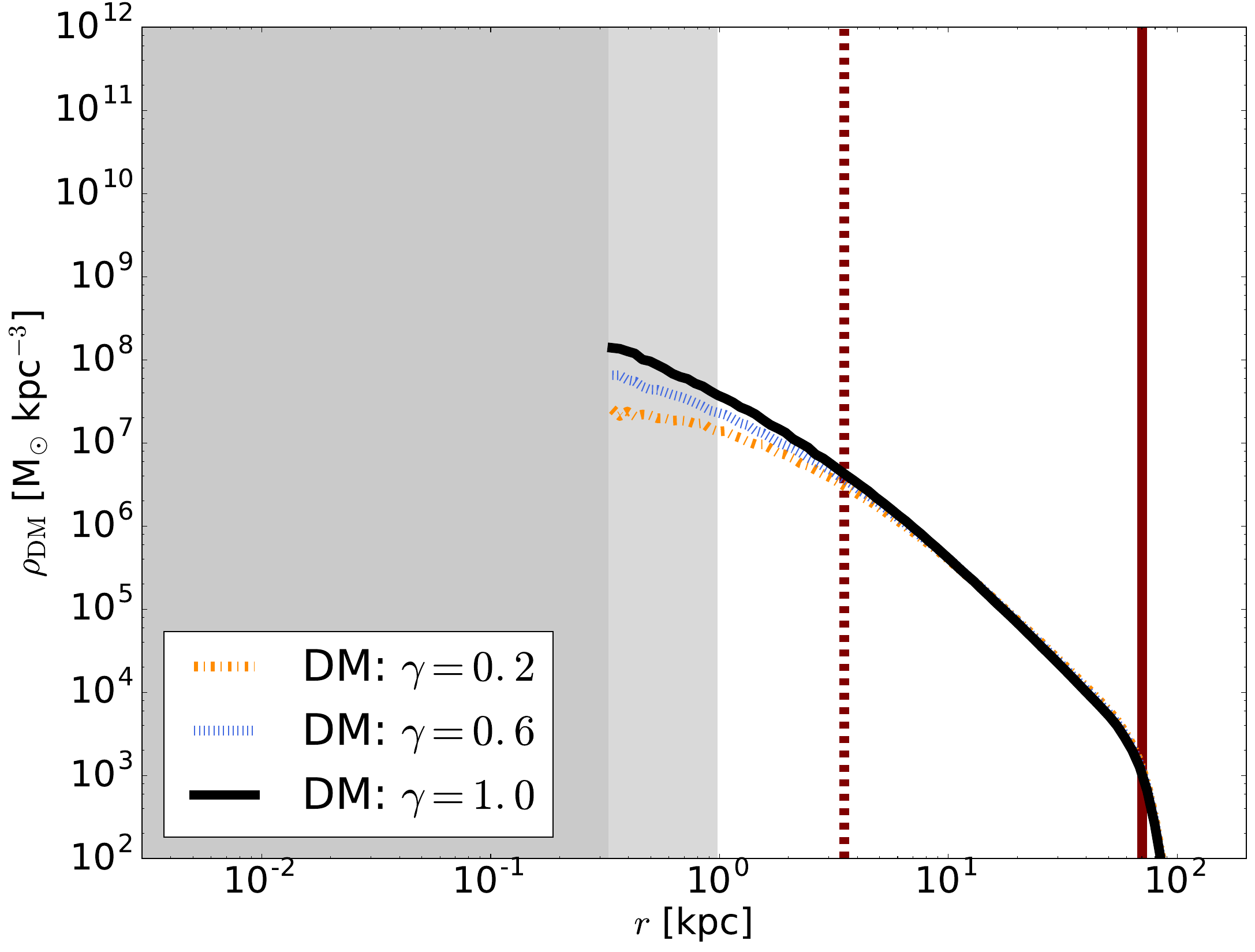}\\
\includegraphics[width=0.48\textwidth]{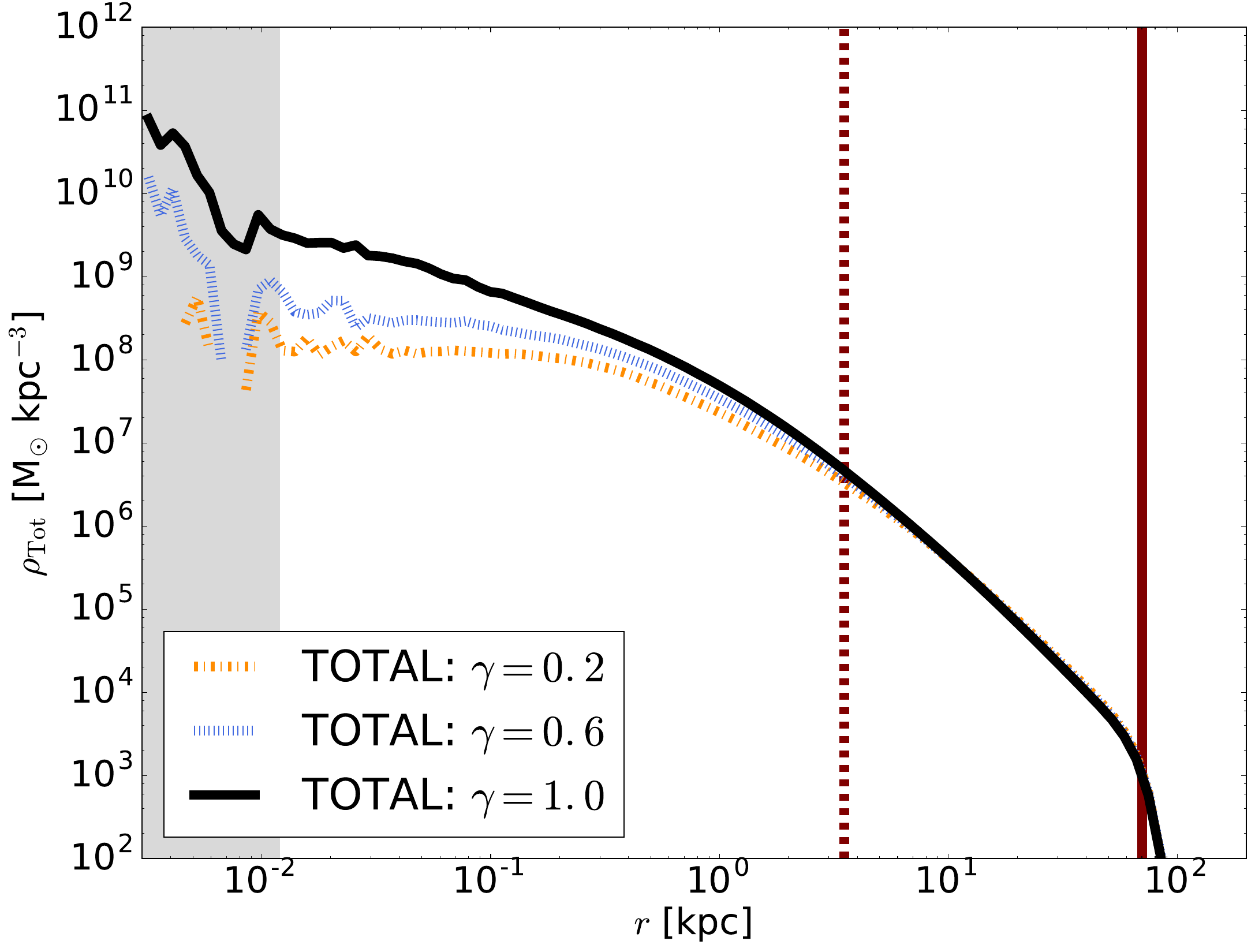}
\includegraphics[width=0.48\textwidth]{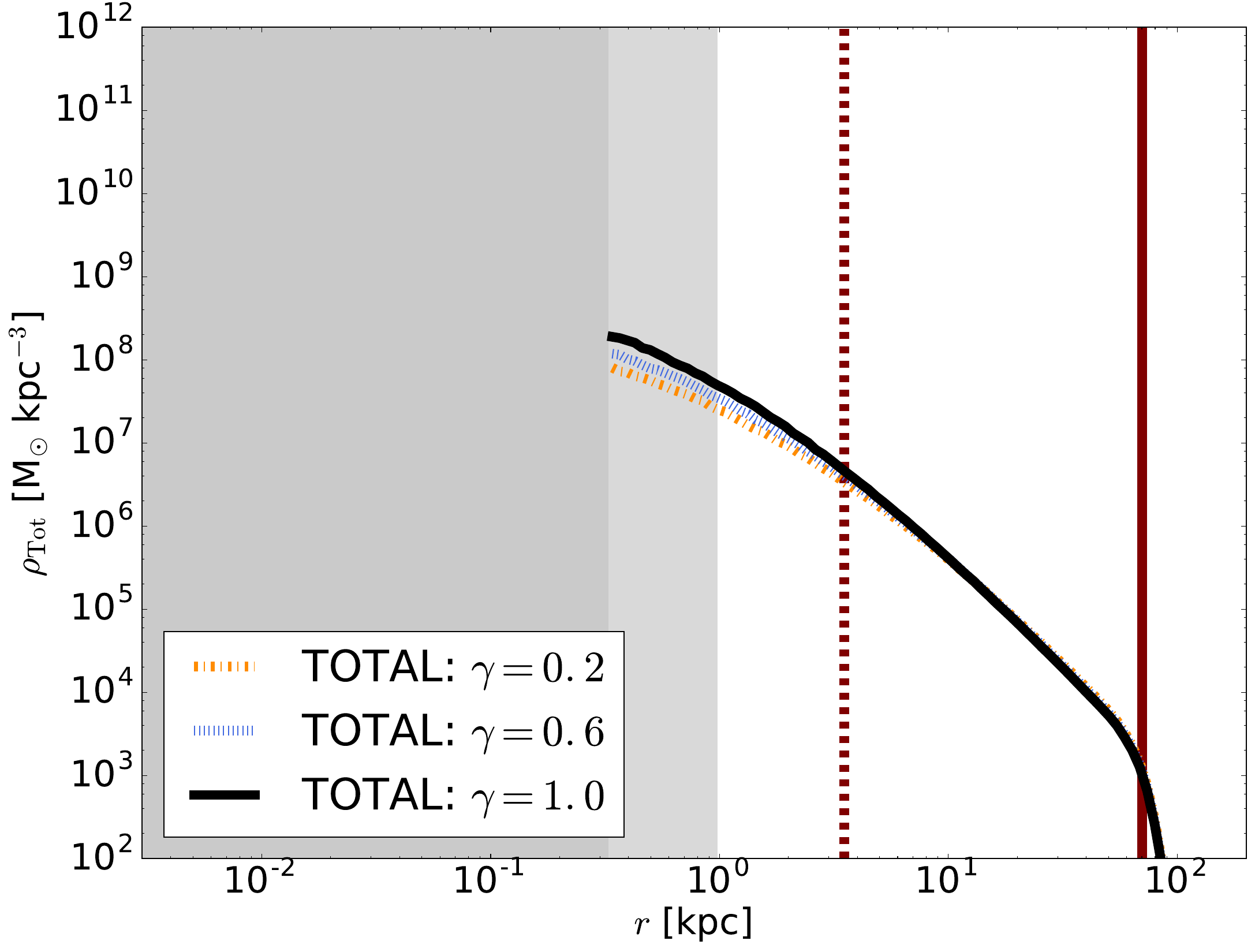}
\vspace{12pt}
\caption{Initial conditions for all merger simulations, run at high (left-hand panels) and low (right-hand) resolution. {\it Upper panels:} Stellar surface-density profiles for three different simulations, $\gamma = 0.2 $ (orange, dash-dotted line), 0.6 (blue, dotted), and 1.0 (black, solid). The vertical line indicates $R_{\rm d}$. {\it Middle panels:} Same as the upper panels, but for the DM volume-density profiles. The vertical lines indicate, from left to right, $r_{\rm s}$ and $r_{\rm vir}$. {\it Lower panels:} Same as the middle panels, but for the total volume-density profiles. {\it All plots:} The dark and light gray-shaded regions represent three times the stellar and DM gravitational softening, respectively.}
\vspace{12pt}
\label{fig:ICFILES}
\end{figure*}

To investigate the degree to which the inner DM density distribution affects the formation of BH binaries during DG mergers, we varied $\gamma$ in three otherwise identically initialized DGs: $\gamma=1$, 0.6, and 0.2 \citep[see also][]{Kazantzidis_et_al_2013}. The value of $\gamma=1$ corresponds to the NFW profile, whereas $\gamma=0.6$ and 0.2 indicate a mild density cusp and a nearly constant density core, respectively. These adopted shallow inner slopes of $\gamma<1$ are well motivated as they resemble those of both observed \citep[e.g.][]{Oh_et_al_2015} and simulated \citep[e.g.][]{Governato_et_al_2010,Tollet_et_al_2016} DGs.

Each DG comprised an exponential stellar disk \textcolor{black}{and a central BH with masses} $M_{\rm d} = 4 \times 10^{8}$~M$_{\odot}$ \textcolor{black}{and $M_{\rm BH} = 10^5$~M$_{\odot}$, respectively}. These values are consistent with the distribution of stellar and BH masses in the DG sample of \citet{Mezcua_et_al_2018}, where they present the largest IMBH sample beyond $z \sim 0$. Moreover, all DG models consisted of a DM halo with a virial mass $M_{\rm vir} = 2 \times 10^{10}$~M$_{\odot}$ (corresponding to a virial radius $r_{\rm vir} = 70$~kpc) and a concentration parameter $c_{\rm vir} \equiv r_{\rm vir}/r_{\rm s} = 20$ \citep[comparable to the median concentration value for a $z=0$ cosmological halo at this mass scale; e.g.][]{Maccio_et_al_2007}. The choice of $M_{\rm vir}$, near the upper limit of the values suggested by empirical models of the stellar mass--halo mass relation \citep[][]{Read_et_al_2017}, was dictated by our desire to employ a relatively low DM-particle-to-BH mass ratio, to prevent numerical two-body heating.

The vertical scale-height and central radial velocity dispersion of the stellar disks were equal to $z_{\rm d} =0.2\, R_{\rm d}$ and $\sigma_{\rm R0} = 20$~km~s$^{-1}$, respectively. Adopting a typical value for the halo spin parameter $\lambda=0.04$ \citep[e.g.][]{Maccio_et_al_2007}, we assigned a disk radial scale-length $R_{\rm d} = 1.1$~kpc \citep{Mo_et_al_1998}\footnote{A critical reader may note that not only adopting a redshift zero for halo formation is at odds with the high-redshift DG mergers relevant to this study, but also employing the \citet{Mo_et_al_1998} formalism may be inappropriate at the scales of DGs. Given the lack of knowledge about the detailed structure of high-redshift DGs and that our scope is to simply determine how the efficiency of BH binary formation in mergers between DGs depends on the inner DM-density distribution, our choices are reasonable and do not bias our results.}, comparable to the effective radius of RGG~118 \citep[][]{Baldassare_et_al_2017}. \textcolor{black}{The above parameters have been chosen according to typical values used in DG simulations \citep[][]{Kazantzidis_et_al_2011} and in forthcoming work a wider range of parameters will be explored.}

The resolution of each component (DM halo, stellar disk, and BH) of the models is listed in Table~\ref{tab:simulations}. We use a typical time-step of 1~Myr, within a hierarchical leap-frog time-stepping scheme which allows time-steps as short as 5~yr. It is important to stress that, with $1.2 \times 10^8$ particles, our high-resolution simulations have a mass resolution nearly two orders of magnitude higher than in recently published cosmological simulations. Each BH was represented by a single particle, placed at the center of the DG initially at rest \citep[][]{Tremaine_et_al_1994}. We evolved all DGs in isolation to assess the adequacy of our numerical choices, confirming that the models are not affected by two-body relaxation and artificial heating of the disk particles via interactions with the more massive DM particles. These test simulations also established that the  residual motion of the BHs around the centers of our DGs does not affect the interpretation of our results, as it is much smaller than the relative separation of the BHs in the cases when a hard binary does not form (see below).

Figure~\ref{fig:ICFILES} presents the density profiles of our DG models after relaxing them in isolation for $0.5$~Gyr, corresponding to ten dynamical times at $1$~kpc.


\subsection{Simulations}\label{sec:simulations}

We generated the initial conditions of the mergers by placing relaxed DGs with the same mass and DM inner slope on parabolic\textcolor{black}{, coplanar, prograde--prograde} orbits \textcolor{black}{(i.e. with the orbital and galactic angular momenta all pointing in the same direction)}, with an initial and first pericentric distance equal to $2\, r_{\rm vir}$ and $0.2\, r_{\rm vir}$, respectively \citep[e.g.][]{Kazantzidis_et_al_2005,VanWassenhove_et_al_2014,Capelo_et_al_2015}. \textcolor{black}{We chose pairs of DGs with the same $\gamma$ because it was numerically shown that the inner DM slope of the remnant, which we will see is crucial for determining if the central BHs will stall, is always close to the steepest slope of the merging galaxies \citep[e.g.][]{Fulton_Barnes_2001,Dehnen_2005}.} We only performed \textcolor{black}{coplanar encounters}, as in those the efficiency of merger-induced torques is maximised, \textcolor{black}{with respect to non-coplanar mergers,} leading to strong stellar redistribution \citep[e.g.][]{Capelo_Dotti_2017} and producing optimal conditions for efficient DF. \textcolor{black}{For the same reason, we expect that, for a fixed initial distance, decreasing the first pericentric distance, hence increasing the efficiency of the torques, would also aid in the formation of a BH pair, although we caution that head-on collisions could end up being too disruptive \citep[e.g.][]{Cox_et_al_2008}.} All merger simulations were carried out with the tree-code \textsc{pkdgrav{3}} \citep{Potter_et_al_2017}. We also performed lower resolution simulations, to study the dependence on resolution of the BH-separation evolution. To be \textcolor{black}{consistent} in all our analysis, we were conservative and chose three times the stellar softening as our limiting factor for the analysis.


\section{Results}\label{sec:Results}

\begin{figure*}[htb!]
\includegraphics[width=\textwidth]{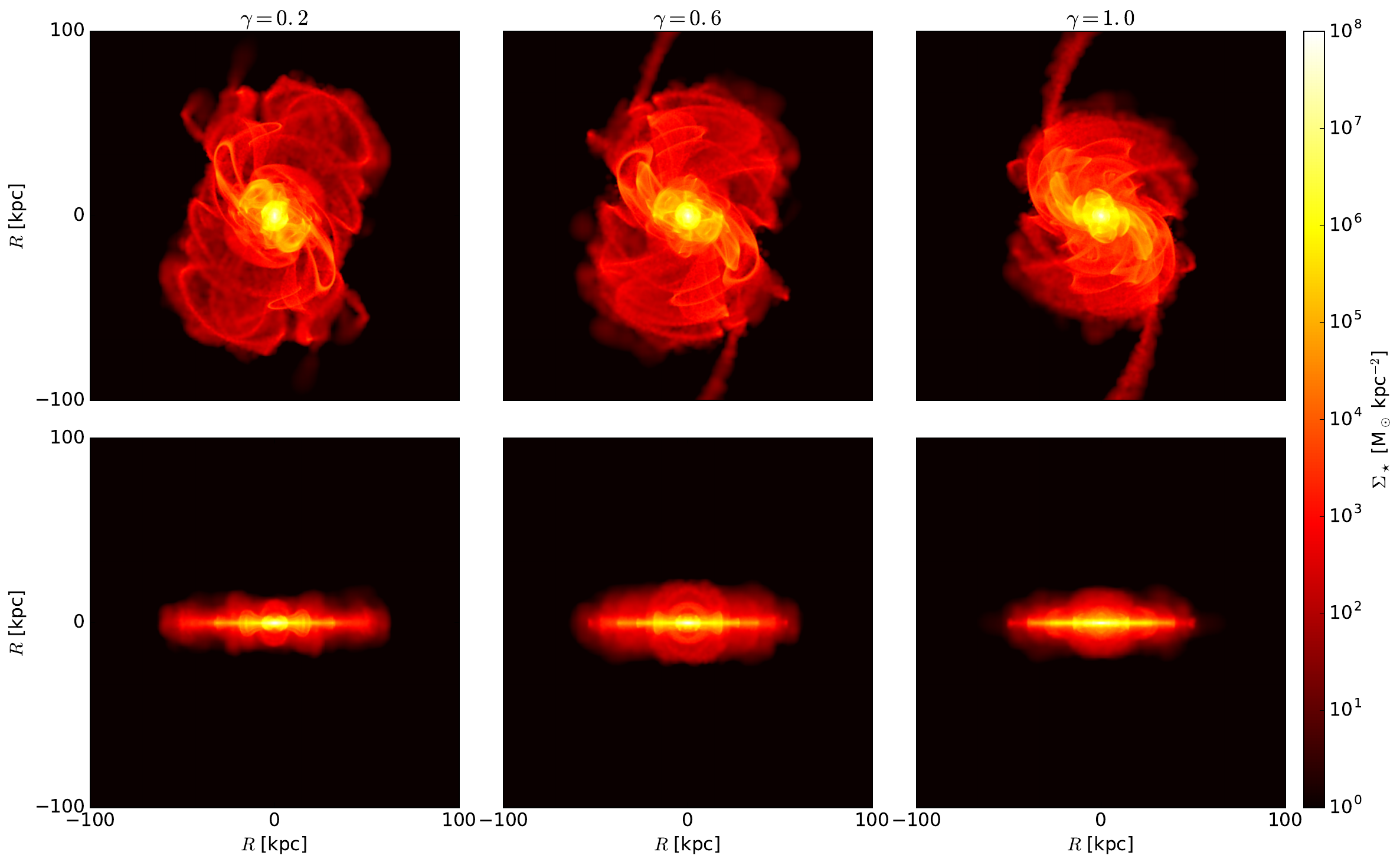}
\caption{Stellar surface-density maps (face-on -- upper panels; and edge-on -- lower panels) of the three high-resolution mergers, shown at $\tau_{\rm Rem}$: $\gamma = 0.2$ (\textcolor{black}{7.43}~Gyr; left-hand panels), 0.6 (\textcolor{black}{7.07}~Gyr; middle), and 1.0 (\textcolor{black}{6.56}~Gyr; right-hand).}
\label{fig:remnantmaps}
\vspace{0pt}
\end{figure*}

Figure~\ref{fig:remnantmaps} presents the stellar surface-density maps of the high-resolution mergers, soon after a remnant has formed, highlighting a qualitative difference between the cuspy ($\gamma = 1$) and shallow ($\gamma < 1$) DGs. The former, having a deeper potential well, produce less extended shells and tidal streams following the merger, resulting in a more compact disky remnant. Tidal tails are also sharper in this case, suggesting that deep photometric observations of the outskirts of interacting DGs might carry information on the underlying DM potential.

\begin{figure}[htb!]
\includegraphics[width=0.47\textwidth]{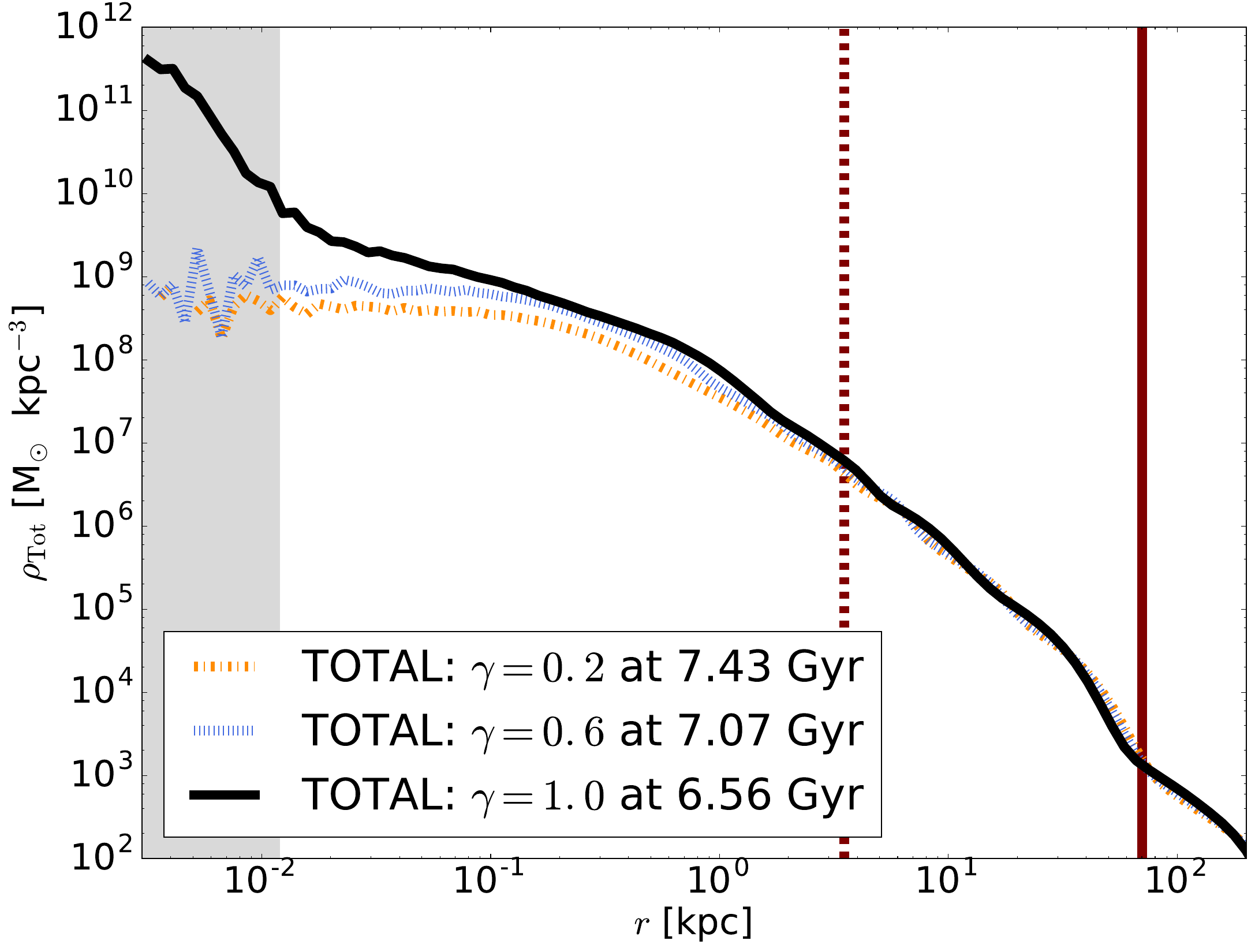}\\
\includegraphics[width=0.47\textwidth]{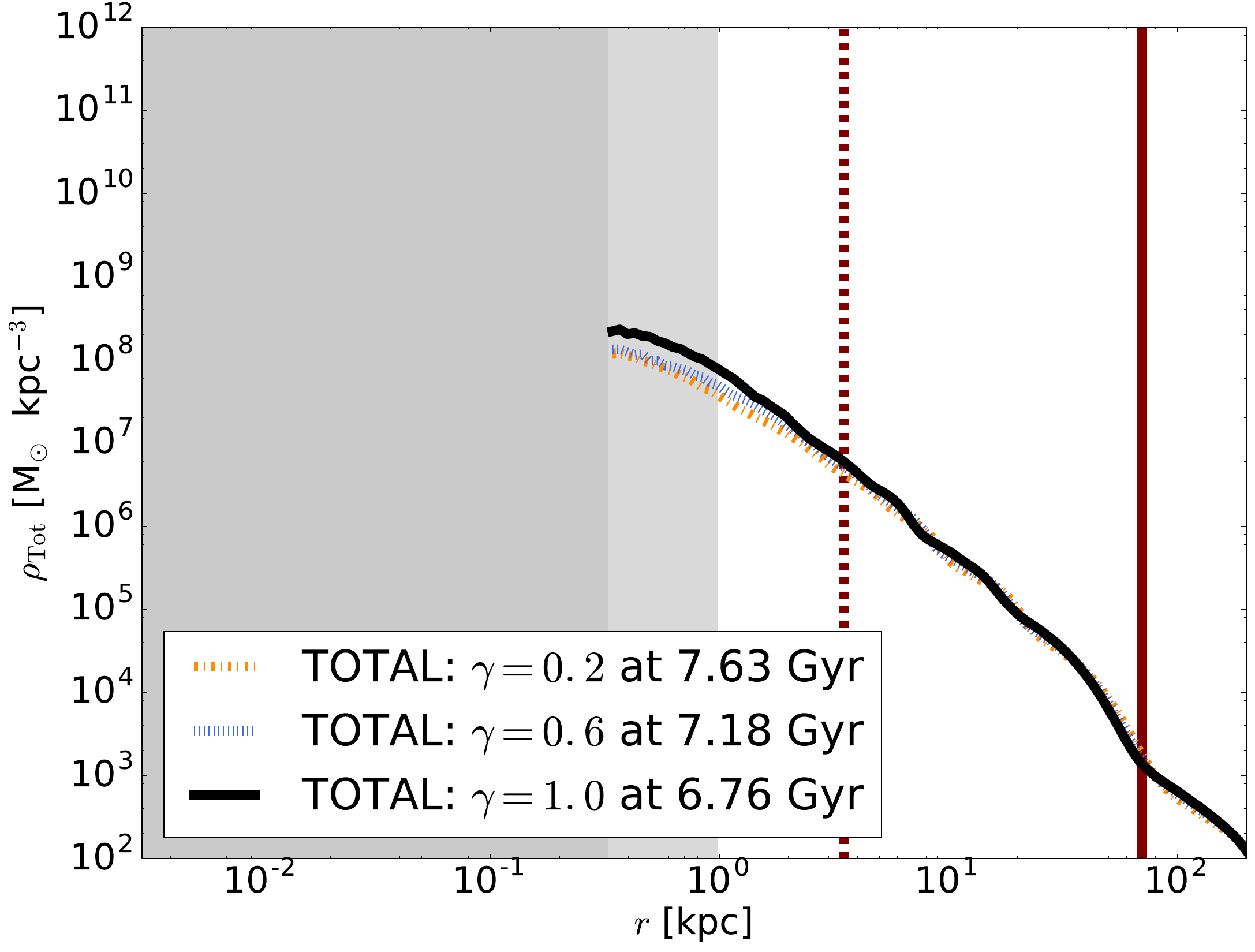}
\caption{Total volume-density profile of the high-resolution (top panel) and low-resolution (bottom panel) mergers at $\tau_{\rm Rem}$. The vertical lines and the gray-shaded regions are the same as in Fig.~\ref{fig:ICFILES}. The central density profiles in the $\gamma = 0.2$ and 0.6 high-resolution cases are flat, whereas the NFW case shows a steeper profile.}
\vspace{0pt}
\label{fig:remnantprofiles}
\end{figure}

The differences among our various experiments are even more striking when comparing the total volume-density profiles of the remnants (Fig.~\ref{fig:remnantprofiles}). In the high-resolution cases, the profiles are nearly flat in the $\gamma <1$ cases for $r \lesssim 0.2$~kpc, whereas the remnant of the cuspy galaxies has a profile even steeper than in the initial conditions in the same region (cf. Fig.~\ref{fig:ICFILES}). These differences arise at radii $\lesssim$1~kpc. Hence, not surprisingly, they are not seen in the low-resolution runs, which appear to have an almost identical slope down to three times the stellar softening.

The structure of the merger remnant has striking implications on the decay of the BHs. In Fig.~\ref{fig:BHsep}, we show the BH-separation evolution for all the simulations. At the beginning, the orbital history of the BHs is, by construction, very similar amongst all runs. However, the different initial $\gamma$ implies a different enclosed mass, which causes the orbits to eventually differ: this can be clearly seen already at the second pericenter and, most importantly, at the respective formation-times of the remnants, highlighted by the vertical lines. Once the BHs are in the merger remnant, stalling occurs at 50--100~pc in the cored models, while the BHs continue to sink efficiently by DF in the cuspy remnant. Such remarkable difference is the key result of this Letter.

We recall that stalling of extended perturbers, such as globular clusters and galaxy satellites, has been widely documented for cored DM halos \citep[e.g.][]{Petts_et_al_2016}. However, here it is significant that the stalling radius is much larger than the distance required for other processes to take over DF and promote decay, as we explain below. In order to show this, we highlight the different orbital-evolution stages of the two BHs by recalling that the BHs will merge on a global timescale $\tau_{\rm Merg}$ given by the sum of three distinct time-scales:

\begin{equation}
\tau_{\rm Merg} = \tau_{\rm Rem} + \tau_{\rm HB} + \tau_{\rm GW}.
\end{equation}

\indent For a fixed mass ratio and orbital configuration (type of orbit, initial separation, and first pericentric separation), the time for the merger to form a remnant, $\tau_{\rm Rem}$, depends on the mass in the central regions of the galaxies. It takes longer to form a remnant in the shallow-slope cases than in the NFW case. To compute $\tau_{\rm Rem}$, we searched for the first occurrence when we do not see two distinct cores in the surface stellar-density maps (Fig.~\ref{fig:remnantmaps}), obtaining, for the high-resolution simulations, $\tau_{\rm Rem} = \textcolor{black}{6.56}$, \textcolor{black}{7.07}, and \textcolor{black}{7.43}~Gyr for the $\gamma = 1$, 0.6, and 0.2 case, respectively.

The second time-scale, $\tau_{\rm HB}$, which is also connected to the central mass distribution and therefore to the density slope of the remnant, is the time that it takes to form a hard BH binary inside the remnant and, in the absence of gas, is driven by three-body encounters with passing stars. Lastly, $\tau_{\rm GW}$ is the time needed by the hard IMBH binary to coalesce via GW emission. The hard-binary separation is given by \citep[][]{Merritt_2013}

\begin{equation}\label{eq:HB}
d_{\rm HB} = \frac{GM_{\rm BH}}{8 \sigma^2},
\end{equation}

\noindent with $\sigma$ denoting the stellar velocity dispersion at 10~pc (computed at $\tau_{\rm Rem}$).

Using Eq.~\eqref{eq:HB}, we obtain $d_{\rm HB} = 0.06$, 0.05, and 0.03~pc for the high-resolution cases $\gamma = 0.2$, 0.6, and 1, respectively. As the stalling radius is much larger than $d_{\rm HB}$ in the cored remnants, it follows that there is no alternative process to promote orbital decay in this case.  Therefore, defining $\tau_{\rm HB}$ by identifying the first time the BH separation reaches $d_{\rm HB}$, it follows that it is indefinitely longer than the Hubble time in the low-$\gamma$ cases, and so is $\tau_{\rm Merg}$ as a result.

We note that the two high-resolution $\gamma<1$ cases behave similarly, as expected from the total density profiles of the remnants (Fig.~\ref{fig:remnantprofiles}). Likewise, small differences are seen amongst all the low-resolution runs (the BH separation oscillates in all cases around 1~kpc, roughly three times the stellar softening) despite the different DM profiles. This is again not surprising, given the small differences in the total matter profiles. Since the high-resolution runs show that the stalling radius is close to $\lesssim$0.1~kpc, it follows that this is not resolved in the low-resolution runs, being comparable to the stellar softening length. Secondly, we verified that the lower number of particles in the low-resolution runs might start to play a role in the late phase of orbital evolution, whereas relaxation time-scales are of order $100\, t_{\rm Hubble}$ in the high-resolution runs, suggesting that our results are robust and reflect the collisionless nature of the systems. Taken together, artificial relaxation and softening considerations cast doubts on studying BH dynamics in cosmological simulations (including most zoom-in runs), since their resolution is usually much coarser than in our high-resolution runs.

The threshold in resolved dynamics induced by gravitational softening  is the reason why, even in the cuspy high-resolution remnant, BHs eventually stop sinking before they can reach the hard-binary stage. However, given sufficient resolution, sinking should continue until the hard-binary stage is reached, likely within $t_{\rm Hubble}$, given the steep orbital-decay curve seen in Fig.~\ref{fig:BHsep}.

In the cuspy case, it is conceivable to  compute $\tau_{\rm GW}$ after the hard-binary stage has been reached. Since we do not use a direct $N$-body code, we can rely on \citet{Sesana_Khan_2015}, who provide an estimate of the coalescence time-scale at the onset of the hard-binary regime. Owing to the low nuclear stellar densities in our remnants, we obtain\textcolor{black}{, for our NFW merger,} $\tau_{\rm GW} \gg t_{\rm Hubble}$, regardless of the orbital eccentricity. However, it is noteworthy to mention that the prototypical DG hosting an IMBH, RGG~118, possesses a compact bulge-like or nuclear star cluster \textcolor{black}{(NSC)} component at its center \citep{Baldassare_et_al_2017}. \textcolor{black}{Moreover, observations of nearby bulgeless DGs with the Hubble Space Telescope (HST) find an association of NSCs with AGN, suggesting that a central density peak is a feature of these systems \citep[e.g.][]{Kormendy_Ho_2013}.}

\textcolor{black}{Thus, in the cuspy case, we can} consider the effect of such central component using \textcolor{black}{for simplicity} a \citet{Hernquist_1990} model with $M_{\rm Bulge} = 0.01 M_{\rm d}$ and $r_{\rm Bulge} = 100$~pc \citep[consistent with the constraints in][]{Baldassare_et_al_2017}, and find $\tau_{\rm GW} \ll 1$~Gyr. The addition of such component \textcolor{black}{(a bulge-like structure or a NSC)} appears thus to be crucial for the coalescence of the BHs in the case in which a hard binary can form. Recalling that IMBH--IMBH mergers should be detectable at least out to $z \sim 10$, hence essentially through the entire cosmic history, BHs inside DGs with cupsy halos should give rise to a significant LISA event rate at IMBH mass scales.


\section{Discussion}\label{sec:Discussion}

Providing quantitative estimates for the number of LISA GW signals from the inspiral and coalescence of IMBHs from DG mergers is extremely difficult, as they depend, amongst other factors, on the BH occupation fraction in DGs, the rate of DG mergers, and the number of DGs.

The AGN occupation fraction in DGs is low, perhaps reflecting the impact of supernova feedback in low-mass galaxies, which removes the gas needed to fuel the central BH.\footnote{There are, however, other possible explanations, such as formation itself and/or dynamics \citep[e.g.][]{Habouzit_et_al_2017,Bellovary_et_al_2018}.} If gas accretion is inefficient, BH mergers then become relatively more important as a BH-mass growth channel relative to larger galaxies, wherein accretion is dominant.

Also, DGs are believed to experience on average three major mergers between $z \sim 12$ and 0 \citep[][]{Fakhouri_et_al_2010}, in contrast to larger numbers when assuming, e.g. Milky Way-sized galaxies (which experience on average five major mergers and several more minor mergers).

While the potentially low BH occupation fraction and the relative rarity of DG merger events might seem detrimental to a high LISA-detection rate, this is outweighed by the fact that DGs with $M_{\rm vir} \sim 10^{10}$~M$_{\odot}$ are $\sim$100 times more numerous at $z = 0$ relative to Milky Way-sized halos and even more so at high redshift \citep[e.g.][]{Reed_et_al_2007}.

However, even with a large integrated number of DG mergers, there remains the question of if (and how fast) a BH binary can form during a DG encounter, which is what we focused on in this Letter.

\begin{figure*}[htb!]
\centering
\includegraphics[width=0.9\textwidth]{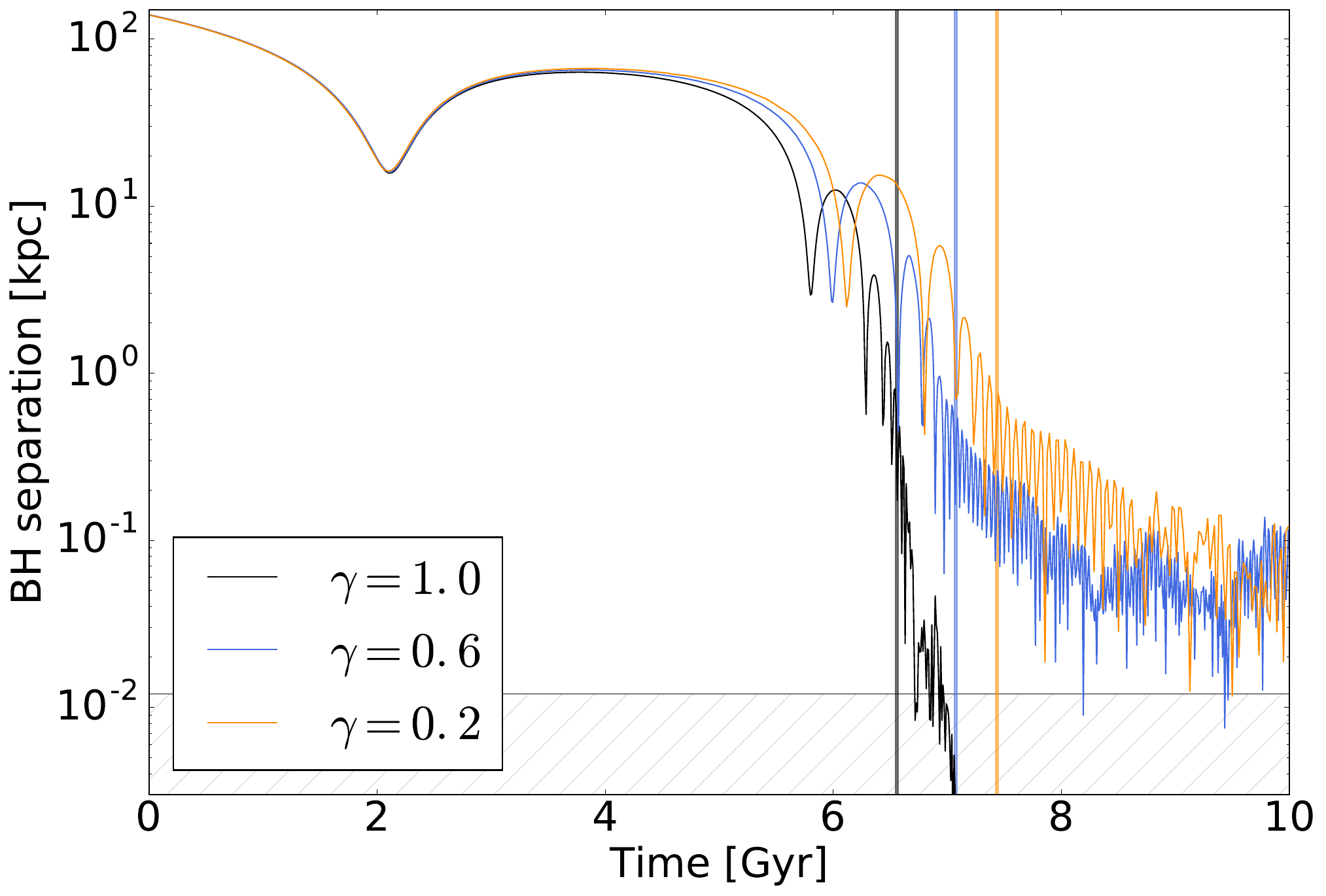}
\includegraphics[width=0.9\textwidth]{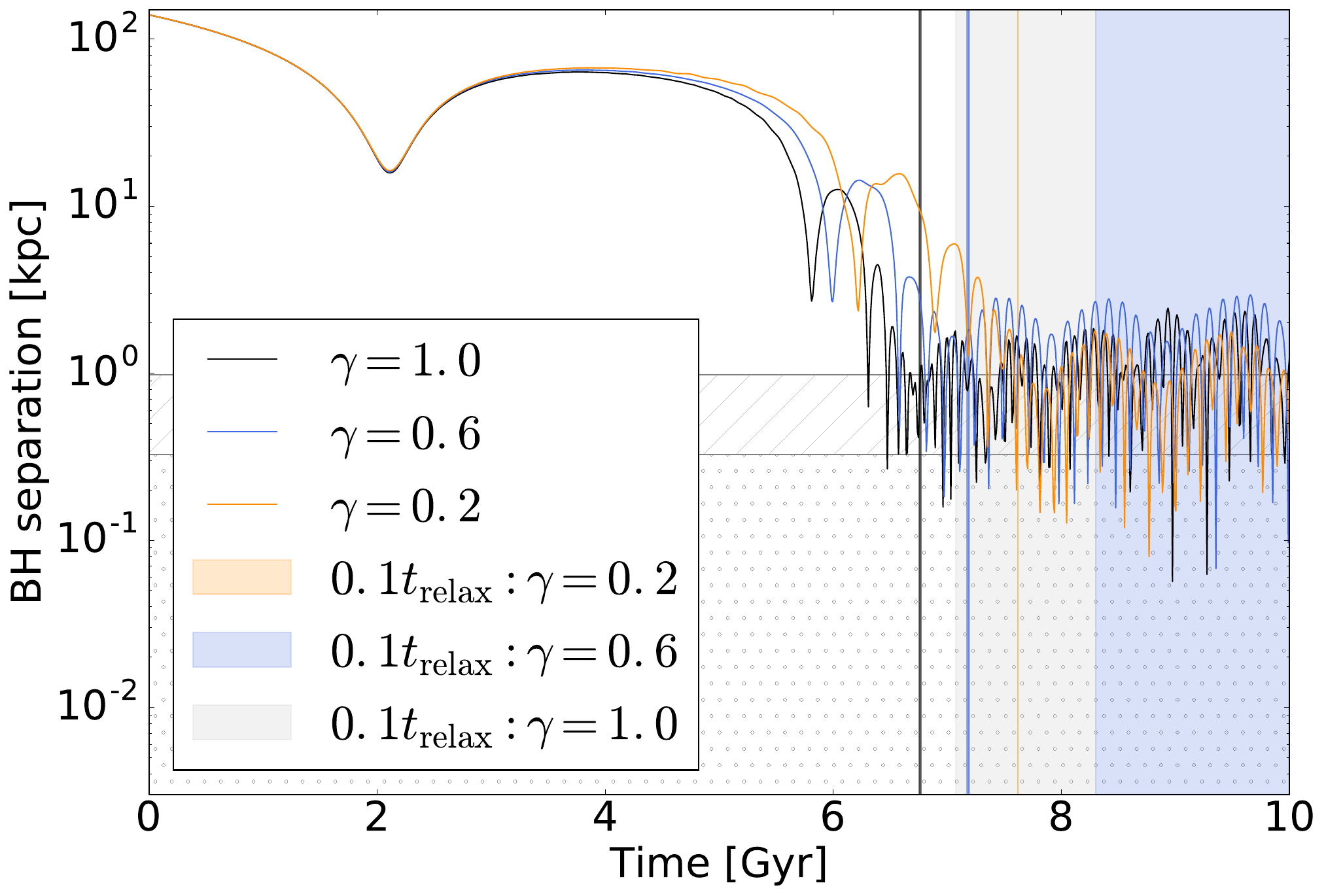}
\caption{IMBH separation versus time for the high-resolution (top panel) and low-resolution (bottom panel) simulations, for the different initial DM slopes $\gamma = 0.2 $ (orange line), 0.6 (blue), and 1.0 (black). The vertical shaded areas indicate 10$\%$ of the relaxation time for the remnant in the inner-kpc region, whereas the vertical lines display the time $\tau_{\rm Rem}$. The horizontal shaded areas show three times the particle softening for stars and DM.}
\vspace{10pt}
\label{fig:BHsep}
\end{figure*}

Our main finding is that the inner DM slope is a fundamental property for predicting the formation (or lack thereof) of an IMBH binary. More specifically, in our high-resolution simulations, only mergers between cuspy DGs ($\gamma$ = 1) favour the formation of a hard BH binary, whereas, in the low-$\gamma$ mergers, the BHs stall at a separation of $\sim$0.1~kpc (Fig.~\ref{fig:BHsep}). As a consequence, the rate of LISA signals from the coalescence of BHs from DG mergers is a function of the fraction of NFW-like DGs. \textcolor{black}{We stress here that, even though we have only performed mergers between DGs with identical $\gamma$, only one NFW-like DG in the pair should suffice, since the inner DM slope of a merger remnant is always close to the steepest slope of the two \citep[e.g.][]{Fulton_Barnes_2001,Dehnen_2005}.} Furthermore, in DM models other than CDM, such as self-interacting or fuzzy DM, in which the formation of constant-density DM cores is a consequence of the physics of the DM candidate, we expect a dearth of LISA events at IMBH scales relative to CDM, hinting at the exciting possibility to use the LISA event rates to probe the nature of DM itself.

Our results depend sensitively on resolution: in the low-resolution simulations, all BHs stall, independent of $\gamma$. This is an important point to keep in mind when interpreting results from cosmological simulations, where the resolution is necessarily limited.

It is obviously very difficult to obtain reliable measurements of the inner DM profile in observed DGs. Moreover, recent hydrodynamic simulations in the CDM cosmogony have shown that a wide variety of inner DM slopes arises owing to baryonic-feedback effects \citep[e.g.][]{Tollet_et_al_2016}, in which case LISA event rates at IMBH scales will probe the physics of galaxy formation at DG scales. Before then, the different stellar distribution arising in remnants with different DM profiles highlighted in  Fig.~\ref{fig:remnantmaps} suggest that wide-field deep photometry is also potentially a probe of the underlying halo structure of DGs. We will investigate quantitatively the latter method in a forthcoming work.

We note that we have included neither a stellar bulge nor a gaseous component in our models. A stellar bulge would in principle increase the central enclosed mass, thereby decreasing the DF time-scale \citep[e.g.][]{Tamburello_et_al_2017,SouzaLima_et_al_2017}. Moreover, the presence of a bulge would also accelerate coalescence, once (if) a hard binary forms. It is, however, currently very difficult to constrain the structural parameters of large samples of DGs in order to provide a quantifiable effect \citep[][]{Mezcua_et_al_2018}.

The addition of a gas disk is indeed critical, as it is not clear if the gaseous component would decrease \citep[e.g.][]{Mayer_et_al_2007} or increase \citep[e.g.][]{Fiacconi_et_al_2013,Tamburello_et_al_2017} the DF time-scale, an issue that in DGs is complicated further by the important effect of supernova feedback on the interstellar medium \citep[e.g.][]{Habouzit_et_al_2017}.

Properly modelling the gaseous component, together with BH accretion and feedback, would be also important in a complementary way: the low-$\gamma$ DG-merger BHs, if they were able to accrete gas, could potentially shine for a long time as dual AGN with separations of $\sim$0.1~kpc. It would therefore be possible to detect them as low-luminosity dual AGN \citep[e.g.][]{Capelo_et_al_2017}, since their distance would be large enough to be resolved by, e.g. \textcolor{black}{HST} (at low redshift). It is actually not excluded that such AGN pairs could be already present\textcolor{black}{, undetected,} in the existing samples \citep[e.g.][]{Mezcua_et_al_2018}.

Given the long stalling time of the BHs in these remnants, it is also possible to have a third body merging with the remnant \citep[][]{Amaro-Seoane_et_al_2010}, with the fate of the triple system depending on the mass of the third body [a possible binary-hardening (ejection), if the third body is less (more) massive]. Given the low average number of mergers experienced by DGs, it is however unlikely that triple DG systems would be common. We defer the investigation of all these issues to future work\textcolor{black}{, together with a detailed analysis of the IMBH--IMBH event rates in the LISA band, since such an analysis would be beyond the scope of this Letter.}


\acknowledgments

We thank Thanos Anestopoulos, Monica Colpi, Massimo Dotti, Zachary Schutte, Nathan Secrest, and Alberto Sesana for fruitful discussions. TT, PRC, SK, and LM acknowledge networking support and SK also acknowledges financial support from the COST Action CA16104, GWverse, on
``Gravitational Waves, Black Holes and Fundamental Physics''. PRC acknowledges support by the Tomalla foundation. \textcolor{black}{LMW was supported by the Natural Sciences and Engineering Research Council of Canada through the Discovery Grant Program.}


\bibliographystyle{aasjournal}
\bibliography{BHbinariesindwarfs}

\end{document}